# An Accurate and Transferable Machine-Learning Interatomic Potential for Silicon

Lin Hu, Rui Su, and Bing Huang*

*Beijing Computational Science Research Center, Beijing 100193, China*

Feng Liu*

*Department of Materials Science and Engineering, University of Utah, Salt Lake City, UT 84112,* and *Collaborative Innovation Center of Quantum Matter, Beijing 100084, China*

[#]Correspondence to: B.H. (bing.huang@csrc.ac.cn) or F.L. (fliu@eng.utah.edu)

# Abstract

The development of modern *ab initio* methods has rapidly increased our understanding of physics, chemistry and materials science. Unfortunately, intensive *ab initio* calculations are intractable for large and complex systems. On the other hand, empirical force fields are less accurate with poor transferability even though they are efficient to handle large and complex systems. Recent development of machine-learning based neural-network (NN) for local atomic environment representation of density functional theory (DFT) has offered a promising solution to this long-standing challenge. Si is one of the most important elements in science and technology, however, an accurate and transferable interatomic potential for Si is still lacking. Here, we develop a generalized NN potential for Si, which correctly predicts the Si(111)-(7x7) ground-state surface reconstruction for the first time and accurately reproduces the DFT results in a wide range of complex Si structures. We envision similar developments will be made for a wide range of materials systems in the near future.

# I. INTRODUCTION

In the past decade, atomistic simulations have become a powerful tool for the study of chemistry, physics and materials sciences. The first-principles based density functional theory (DFT), with the only input of atomic elements, has shown remarkable success in solving a wide range of problems in materials science as well as predicting new previously-unknown materials. However, there remains a long-standing challenge that the accuracy of DFT is at the expense of huge computing cost. In general, it is limited to simulating a system containing hundreds of atoms for a time scale of hundreds of picoseconds or even nanoseconds based on DFT. In addition, the algorithmic scaling of common DFT methods is overall super linear with the fastest portion at $O(NlnN)$, where $N$ is the number of atoms in the system. This means that the computation may suffer a poor scaling with the increasing size of the simulated system.

The empirical-potentials calculations, with a parameterized Born-Oppenheimer potential-energy surface (PES), have been widely adopted to accelerate the computational speed compared to the DFT calculations [1-7]. Differing from the DFT calculations, the algorithmic scaling based on empirical potentials is almost linear with the number of atoms. However, empirical potentials rely on a small number of parameters that are adjusted to reproduce either existing experimental or first-principles references. Therefore, it is not surprised that the empirical potentials are not sufficiently accurate and transferable, i.e., it cannot be generally adopted to describe many different properties of a system simultaneously.

In order to resolve the above-mentioned problem, recently, a number of methods have been proposed by employing machine-learning (ML) algorithms to fit the first-principles PES in high dimensional spaces instead of using simple parameterized functional forms [8]. Among them, the most popular methods are by Behler and Parrinello (BP) which are based on artificial neural networks (NN) [9,10] and Gaussian Approximation Potential (GAP) [11,12]. Both methods require the design of different routines to accurately describe the local structural environment of atoms and using basic functions, e.g. Gaussian functions, to construct the potential, then the parameters are fit with either linear algebra or some other protocols of NN optimization [13,14]. It is noted that there is no specific physical model but only a pure mathematical expression with a large number of adjustable parameters in the ML potentials that can be done automatically without much manual intervention. Since the input of NN potential depends solely on the local atomic environments, the targeted systems of the same/similar local environments with different sizes can be well described and predicted. Consequently, one can use small supercell structures to create training databases and then obtain accurate predictions for large-scale systems with similar local atomic environments. Ideally, the NN potential can achieve the similar accuracy as the first-principles DFT method.

However, there are some disadvantages for ML-based NN potentials. The Achilles heel of ML methods is that these high-dimensional fits are usually good for interpolation rather than extrapolation. The transferability and accuracy of ML potentials are mainly determined by the selected training database, i.e., it is difficult to accurately describe the physical properties of those structures with large diversities beyond the database. Thus, in order to obtain a general potential, it is necessary to create a sufficiently large training database to cover as many structural information of the materials as possible.

Silicon is one of the most important elements in our daily life and industry. Fundamentally, Si is also often chosen as the prototype materials system to test new computational method [15-17], demonstrate new experimental approaches [18-21] and illustrate new theoretical concepts [22-25]. However, a general and accurate Si interatomic potential is still lacking, especially there is no interatomic potential available to correctly describe the complicated surface reconstructions of Si [22]. In this work, a generalized NN potential for Si is obtained, which can accurately reproduce the DFT results for a wide range of properties, including bulk crystals, liquid, amorphous systems, point defects, and surface reconstructions. Most remarkably, our NN potential is the first and only one that can predict and describe the ground-state dimer-adatom-stacking-fault (DAS) reconstruction of Si(111)-(7x7) surface [18,22]. It is expected our NN potential could be applied as one of benchmarks for the future study of complex Si structures.

## II. METHODOLOGY

### A. Principle

The basic idea of NN potential is to automatically determine and parameterize the model that can describe the featured space of input data, instead of fitting a predefined model or function, with a certain training database or existing reference data. The potential fitting process that learns the relationship between atomic structure and energy from the training database is called *supervised learning* model. Once the model is successfully trained, it can be used as a *black box* to predict various physical properties of materials, e.g., energy and atomic force. It is noted that special attentions need to be paid to the transferability of ML potential.

The NN computing algorithm is inspired by the graphical representation of our brain's NN. The neurons are connected into a network to receive or transmit electrical signals when the input signal exceeds a certain threshold. This analogy leads to the design of the NN potential, albeit signal activation is expressed by a suite of mathematical functions. As the generalized NN potential adopted in this study, BP method [9,10] is one kind of feedforward NN [26] that contains three parts, i.e., *input layer*, *hidden layer* and *output layer*. In each hidden layer, between the input and output layers, there should be sufficient nodes, similar to the neurons in our brain [9]. In the nodes of input layer, a set of *m* symmetry functions related to the atomic coordinates $G_i^1 \ldots G_i^m$ are provided, and the node in the output layer will attain the corresponding energy $E_i$. All the nodes in each hidden layer are connected to one node in the adjacent layers, and the output from the former layer will become the input of the next layer with adjustable weight parameters. The weight parameters are initially randomly chosen, which will be optimized to fit the PES after training. For example, for a simple one-hidden-layer (*n* nodes) NN, the output layer energy can be written as:

$$E_i = f_a^2[w_{0i}^2 + \sum_{j=1}^{n} w_{ji}^2 f_a^1(w_{0i}^1 + \sum_{k=1}^{m} w_{kji}^1 G_i^k)] \,. \tag{1}$$

Here $w^1_{kji}$ is the weight parameter of connecting node $k$ in input layer with node $j$ in hidden layer, and $w^2_{ji}$ is the weight parameter of connecting node $j$ in hidden layer with $E_i$ in output layer. $w^1_{0i}$, $w^2_{0i}$ are biased weight parameters that can be used to adjust the offset of the activation functions $f^1_a$, $f^2_a$. In principle the activation function should be a step-like signal function, but in practice the hyperbolic tangent function ($tanh(x) = \frac{1-e^{-2x}}{1+e^{-2x}}$) or hyperbolic tangent function with linear twisting function ($1.7159\, tanh\left(\frac{2x}{3}\right) + ax$) are recommended for general purposes [26]. Some ML methods are used to optimize the weight parameters to minimize the errors between input and output (based on DFT total-energy calculations) iteratively.

The first step to construct a NN potential is to transform the atomic coordinates into a set of symmetry functions $\{G^k_i\}$. These symmetry functions must ensure that the total energy is invariant to interchanging of two atoms with the same kind of element so that the potential is flexible for system with arbitrary size. Furthermore, a crucial point is to include a suitable symmetry function that also can describe the local chemical environment around each atom in the system. To satisfy this requirement, the symmetry functions are often constructed as the empirical potential with atomic coordinates. To define the relevance of local chemical environment around each atom, a cutoff function $f_c$ is used as the following:

$$f_c(R_{ij}) = \begin{cases} 0.5 \times \left[\cos\left(\frac{\pi R_{ij}}{R_c}\right) + 1\right], & \text{for } R_{ij} \leq R_c, \\ 0 & , \text{for } R_{ij} \leq R_c. \end{cases} \quad (2)$$

Here, the $R_{ij}$ is defined as the cutoff radius, which should be set sufficiently large to include all nearest neighbors. To preserve the invariance of total energy with respect to translation and rotation, two forms of symmetry functions should be taken into account: radial and angular symmetry functions, which are respectively expressed as:

$$G^1_i = \sum_{j \neq i}^{all} e^{-\eta(R_{ij} - R_s)^2} f_c(R_{ij}). \quad (3)$$

$$G^2_i = 2^{1-\zeta} \sum_{j,k \neq i}^{all} (1 + \lambda \cos\theta_{ijk})^\zeta \times e^{-\eta(R^2_{ij} + R^2_{ik} + R^2_{jk})} f_c(R_{ij}) f_c(R_{ik}) f_c(R_{jk}). \quad (4)$$

For the radial part, $\eta$ and $R_s$ are constants, $\vec{R_{ij}} = \vec{R_i} - \vec{R_j}$. For the angular part, $\theta_{ijk} = \frac{\vec{R_{ij}} \cdot \vec{R_{jk}}}{R_{ij} R_{jk}}$ is constructed to describe three-body interactions. Here, Gaussian function is used to ensure a smooth decay to zero when the interatomic distance is very large. Beside Gaussian function, other symmetry functions can also be used to describe the local atomic environment [27].

Three kinds of training methods have been used for the weight optimization in the NN potential: (1) *Gradient Descent (GD)*. *GD* is a simple method to iteratively minimize the gradient of error function, but it cannot be well parallelized. (2) *Limited-memory BFGS (L-BFGS)*. *L-BFGS* is often regarded as a specific form of the *quasi-Newton* method in NN potential training process, which can be well parallelized. (3) *Levenberg-Marquardt (LM)*. *LM* method is another standard method based on least-square-fitting and it is also very efficient when the NN architecture is very small. In this work, *L-BFGS* is adopted for our NN potential training.

**B. Database**

Database is an important component of NN potential, which is critical to define the accuracy of NN potential. In order to accurately describe or predict a large number of systems, a sufficiently large number of atomic configurations (>10000) of different Si systems are included in the training database. The choice of the database is to balance the computational cost and the transferability of the NN potential. In principle, a more general NN potential needs a larger database, but the model will become more complex and the computational cost of training will also be greatly increased. We use many different silicon systems, including bulk crystals, amorphous, point defects and surface structures, to represent the diversity of local atomic environments. The configurations were chosen by intuition and experience, combined with feedback from training and testing results.

All the first-principles calculations were carried out employing the plane augmented wave (PAW) method as implemented in the first-principles DFT package VASP [28,29]. The generalized gradient approximation (GGA) within the framework of Perdew-Burke-Ernzerhof (PBE) functional is adopted for the electron exchange and correlation [30]. An energy cutoff of 520 eV is employed for the plane wave. Monkhorst-Pack k-point grids are with 0.05 Å$^{-1}$ spacing [31], and 0.01 eV smearing of band filling.

## C. Training and testing

Differing from the empirical potential, there is no explicit physical meaning in the functional form of the NN potential. In addition to stable structures, we emphasize that the structures used for training the potential also include many metastable phases under different pressure and temperatures (simulated by MD calculations). Tuning the NN architectures, i.e., numbers of hidden layers and nodes per hidden layer, a good fitting of NN potential can be obtained to perform MD or Monte Carlo (MC) simulations. The root-mean-square-error (RMSE) method is widely accepted to judge the quality of fitting. Whenever the RMSE is larger than the error of fitting, additional DFT reference data needs to be included into the database until the best fitting is achieved after training.

It is noted that the testing database is chosen depending on training database. If the DFT reference database is large enough, there is a simple but effective way to generate testing database, which is called simple cross validation (SCV). In SCV, the data set is randomly divided into two parts: training set and testing set. The training set is used for training and the testing set is used to do the final evaluation of the learning method.

For our NN potential, about 14000 DFT atomic configurations, containing between 8 (bulk Si) and 300 [Si(111)-(5×5) DAS reconstructed surface] atoms, were used as the reference database, 12600 (90%) of which were used for training and optimizing the NN potential and the rest 1400 (10%) were randomly chosen to test the transferability of the NN potential, which is sufficient to determine under- or over-fitting. Usually, we choose 2 hidden layers with 10 nodes. In the 32 Gaussian symmetry functions, 16 radial and 16 angular symmetry functions have been used as the input nodes. The RMSE of the optimization (testing) set is about 3.9-4.1 (4.0-4.1) meV per atom. We have also made a full test on the convergence of different number of nodes ($N$) in hidden layer and found that $N$=10 is sufficient to achieve a good fitting with this database.

## III. RESULTS

In this section, we report the comprehensive calculations using our developed NN potential for various Si systems, focusing on a large number of different physical properties. We compare our NN potential with the existing empirical potentials to demonstrate the advantages of the NN potential in terms of transferability and accuracy. A large number of empirical potentials have been developed for Si, and the widely used ones are Stillinger and Weber (SW) [32,33] and Tersoff [34-38] potentials. Both SW and Tersoff potentials include the pair and three-body terms that are fit to limited bulk Si properties. Many efforts have been made to improve these potentials for better transferability, e.g. environment dependent interatomic potential (EDIP) [39], modified embedded atom method [40,41], ReaxFF [42] and screened Tersoff [43,44]. In addition, we have compared our results with other complex empirical potential models, density-functional tight-binding (DFTB) method [45-47] and the recent ML-based GAP potential [48].

Below, we will demonstrate the accuracy of our Si NN potential by the comparison with the existing popular potentials or methods. In subsection A, we focus on the total energy of different Si bulk phases and the elastic properties of diamond Si. In subsection B, we focus on the melting point calculated using NN potential. In subsection C, we focus on amorphous Si structure. In subsection D, we focus on different point defects in Si. In subsection E, we focus on Si surface reconstructions calculated using NN potential.

### A. Bulk crystals

Firstly, we have calculated the total energy as a function of hydrostatic strain for six different crystalline Si structures, including diamond (dia), body centered cubic (bcc), high pressure structures (bct and hp), face centered cubic (fcc), and hexagonal diamond (hex-dia). Figure 1 shows the calculated energy-strain curves, together with the relative errors compared to the DFT results. Generally, there are excellent agreements between our NN potential and DFT for the calculations of bulk modulus and the equation of state.

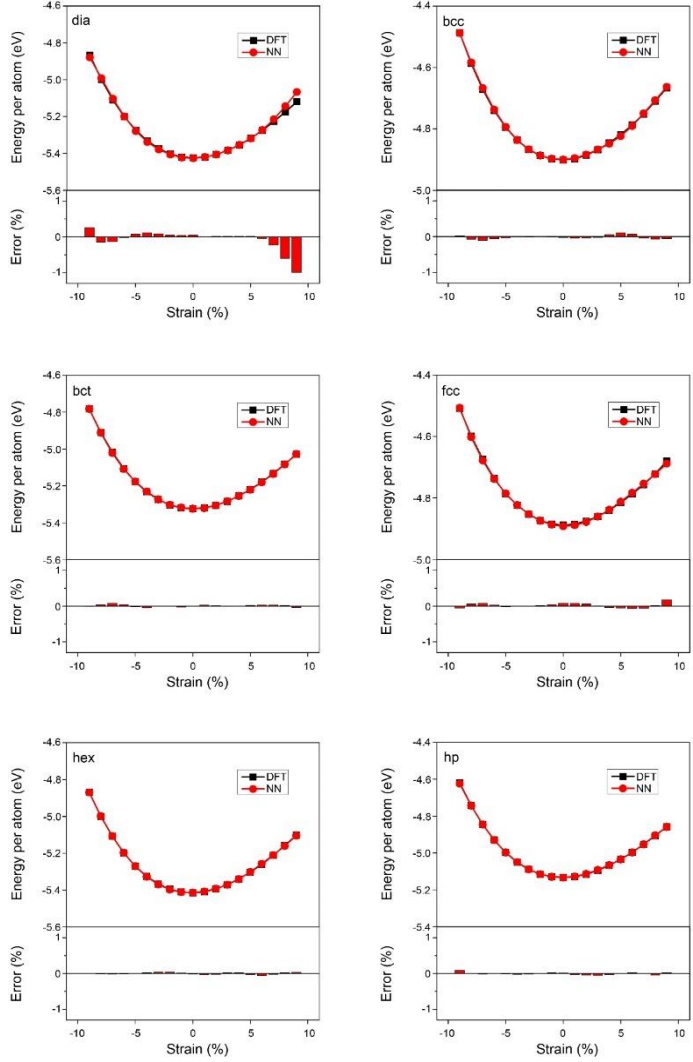

FIG. 1. Energy per atom as a function of hydrostatic strain calculated using NN potential (red line) and DFT method (black line) for various bulk Si crystal structures. Percentage errors are made with respect to DFT reference.

In addition, we have compared our NN potential calculations to other empirical potentials calculations. For bulk properties, we have calculated bulk modulus $B$ and elastic constants $c_{11}$, $c_{12}$ and $c_{44}$ with the Si diamond structure. As shown in Table 1 and Fig. 2, the relative errors of NN potential are less than 10% compared to the DFT, similar to the ML GAP potential and much smaller than other empirical potentials [48].

TABLE 1. Bulk modulus $B$ and elastic constants $c_{11}$, $c_{12}$ and $c_{44}$ calculated by DFT and NN potential.

| Model | $B$/GPa | $c_{11}$/GPa | $c_{12}$/GPa | $c_{44}$/GPa |
|---|---|---|---|---|
| DFT | 91.4 | 153.0 | 55.8 | 74.5 |
| NN potential | 90.6 | 151.9 | 54.1 | 78.3 |

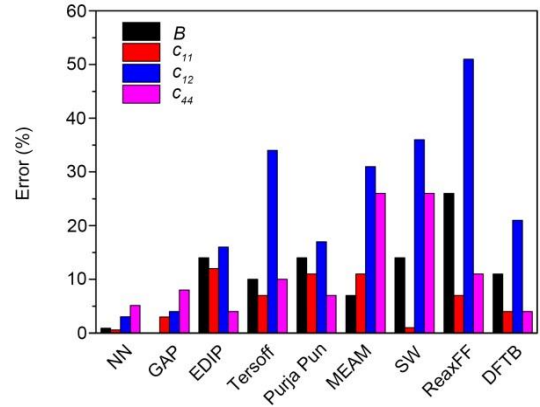

FIG. 2. Comparison of percentage errors of bulk modulus $B$ and elastic constants $c_{11}$, $c_{12}$ and $c_{44}$ calculated by NN potential, GAP model, a range of empirical potentials and DFTB method [45].

## B. Liquid

An accurate prediction of the melting point of Si using first-principles methods is very challenging, because the free energy needs to converge with system size and simulation time, which requires very high precision [49]. Although using molecular dynamics (MD) simulation with empirical potentials can reduce computational time, a large error ~50% can exist in the melting point compared to the experimental value [37,50].

To demonstrate the ability of our NN potential on extracting the melting temperature of diamond Si, a supercell of 5832 atoms was heated in the canonical ensemble (NVT) in 100,000 steps using a 2 fs time step with LAMMPS software [51,52]. Periodic boundary condition (PBC) is implemented. As shown in Fig. 3, the melting temperature is calculated to be 1630 K using Lindemann criterion [53], which is slightly lower (by ~57K) than the experimental melting temperature of 1687 K [54]. For comparison, the melting point predicted by DFT methods with GGA (hybrid functions) is ~200 (~150) K lower (higher) than the experimental value [55-57]. Our result is comparable to the DFT result with random phase approximation, which predicts the melting temperature of 1735 and 1640 K without and with core polarization effects of the electrons, respectively [49]. The melting temperature calculated by our general NN potential is better than the one obtained from the modified Tersoff-ARK potential (1616 K), which is a specially targeted for predicting the melting point of Si [58].

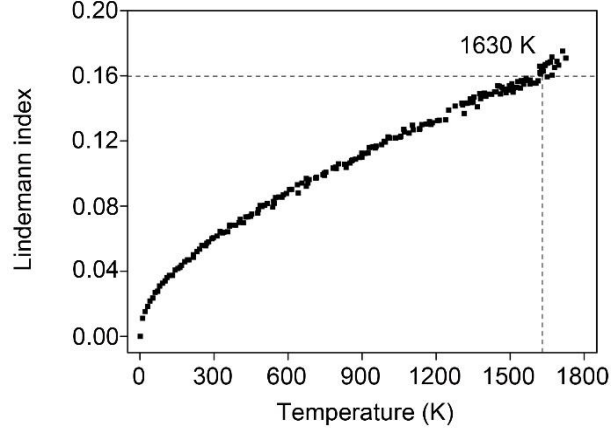

FIG. 3. Variation of the Lindemann index upon heating.

## C. Amorphous phase

In the crystalline Si, Si atoms are fourfold-coordinated and tetrahedrally bonded to the neighboring atoms. In an amorphous Si (a-Si) structure, there are no such long-range ordered tetrahedral bonds, i.e., Si atoms form a continuous random network (CRN) with over- or under-coordination. The dangling bonds associated with the under-coordinated Si in a CRN may result in abnormal electronic behavior. At 300K, the density of amorphous Si is estimated to be $4.90 \times 10^{22}$ atoms/cm$^3$ (2.285 g/cm$^3$) in the experiment [59].

There are two typical approaches to generate amorphous structures. One is to generate a CRN, which can be constructed using a variety of methods to achieve a structure with only 4-fold coordinated atoms. Then, using a reverse Monte Carlo (RMC) simulation, a three-dimensional particle configuration can be generated to fit the experimental structural factors [60]. The other method is to simulate rapid quenching of liquids. However, the conventional empirical potentials cannot accurately describe the high-density structural defects in the bulk phase, giving rise to inconsistent structural information of a-Si compared to the experimental measurements. Although *ab initio* MD [61] can accurately describe electronic interactions, due to limitations of computational cost, it can only treat small systems (~100 atoms) with very short time scale (~10 ps), which may lead to systematic errors in the simulation. Also, such a short simulation time can result in a huge cooling rate, which is far from the experimental condition.

In this work, we have performed the simulations of quenching several Si supercell (512 atoms) samples to obtain the liquid Si using a similar simulation process in subsection B. The sample is cooled down at $10^{12}$ K/s from 2000 K to 300 K, in accordance with the laser experiments [62].

The RDF is widely used to describe the structural characteristics of an amorphous system. Fig. 4 shows the calculated RDF using NN potential, in comparison with the experimental results [Fig. 4(a)] and RMC simulations [Fig. 4(b)) [60]. The results obtained from our NN potential agree well with the experimental and MC simulations. The first-neighbor peak is located at ~2.35 Å and the relatively broad second peak sits at around 3.80 Å. It is worth noting that most other potentials, e.g., GAP, EDIP and Tersoff potential, produce a zero-atom-distribution in the range from 2.5 Å to 3.2 Å, in disagreement with the experiments [48]. Therefore, our NN potential provides arguably the best interatomic potential to date for simulating the structures of amorphous Si.

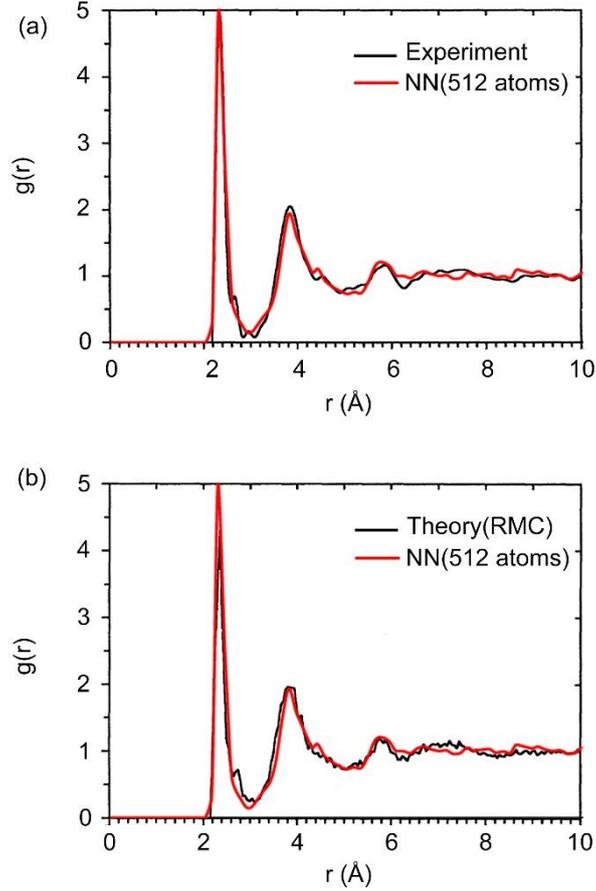

FIG. 4. Radial distribution function (RDF) for 512 atom amorphous Si configuration. For NN (red line) model in both (a) and (b). Experimental result (black line) in (a) and reverse Monte Carlo (RMC) simulation (black line) in (b) are from Ref. [60] for comparison.

### D. Defect

Point defects in bulk Si have been widely studied in the past decades. There are three types of point defects in Si: vacancies, interstitials and Frenkel pairs. For interstitials, there are two types: one is (110) $X$ interstitial and the other is hexagonal $H$ interstitial. The total energy per atom as a function of tri-axial strain is calculated using DFT and NN potential, as shown in Fig. 5. The overall agreement between the two methods is excellent, i.e., all the errors are within 1%. It is noted that these point defects are not directly included in our database during the training process, indicating a very high degree of transferability of our NN potential.

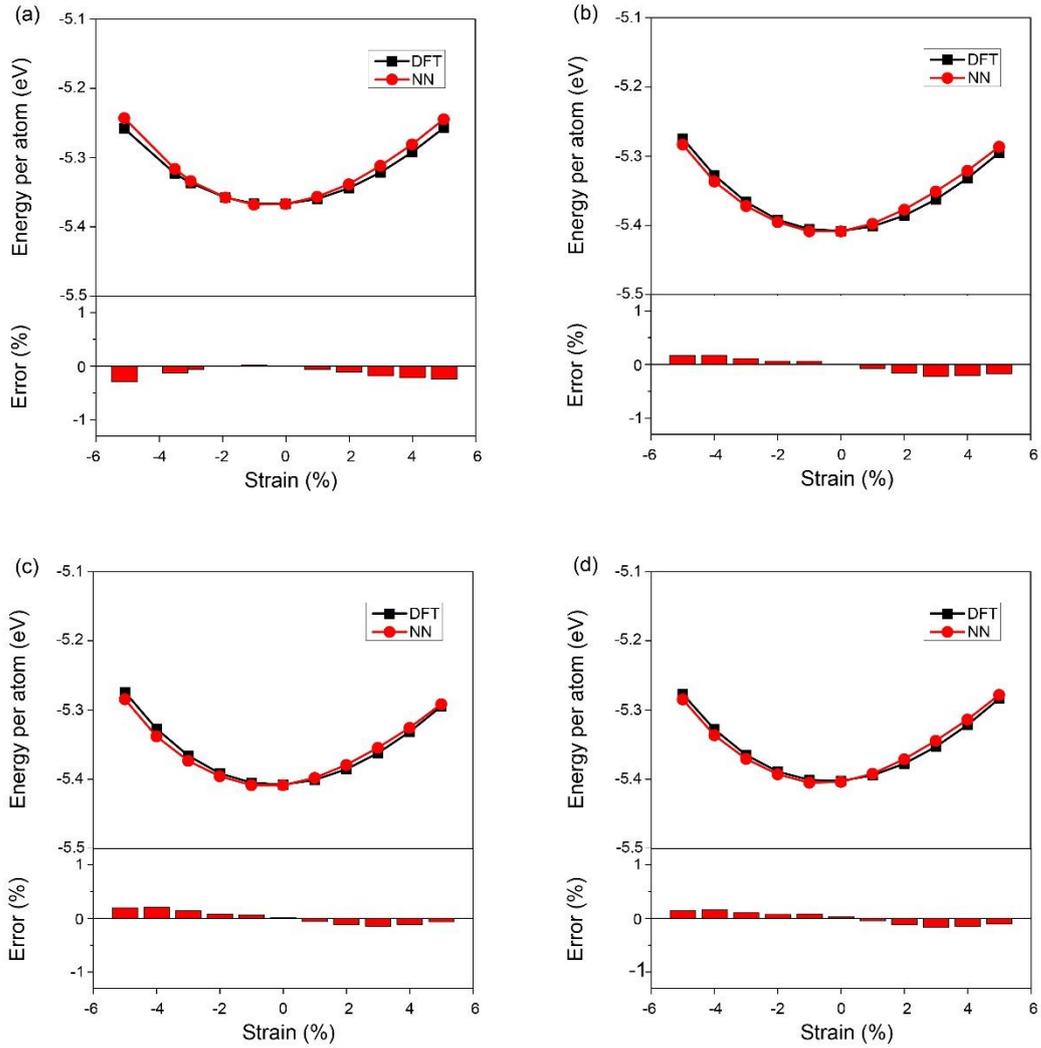

FIG. 5. Energy per atom as a function of tri-axial strain calculated using NN potential (red line) and DFT (black line). The comparison of percentage errors between NN potential and DFT reference for (a)vacancy, (b)*X* interstitial, (c)*H* interstitial and (d)Frenkel pairs.

We have also compared the relative errors for vacancy, two interstitials and Frenkel pairs between our NN potential with GAP and other empirical potentials [48], as shown in Fig. 6. Our NN potential shows much smaller errors than others, including GAP.

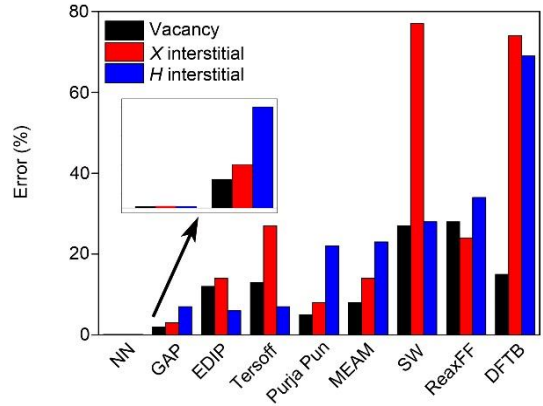

FIG. 6. Comparison of percentage errors of vacancy, *X* interstitial and *H* interstitial calculated by NN potential, GAP model, a range of empirical potentials and DFTB method [48]. Insert is a zoom-in for NN potential and GAP method.

### E. Surface

Surface plays a critical role for modulating the electronic properties of a semiconductor. The chemical reactions mostly occur on the surface. Surface structure also plays a key role in epitaxial growth of thin films [20-25,63-66]. Due to the subtle competition between the strain effect and the chemical effect of dangling bonds, a rich bonding complexity occurs on the surface. Consequently, it is very challenging to accurately describe the surface structures using interatomic potentials.

Basically, surface reconstruction can remove dangling bonds in the surface, which can lower surface energy. On the other hand, surface reconstruction changes the interatomic spacing and bond angle, which causes surface strain that increases surface energy [22]. In addition, strain can be relaxed by changing surface morphology via formation of surface steps and dislocations [20-21]. The strain relaxation process is always accompanied by a redistribution of the surface stress field, which is often accompanied with reconstruction. For Si, there are two most important reconstructed surfaces: (001) and (111) surfaces [67].

#### 1. Si(001) reconstructed surface

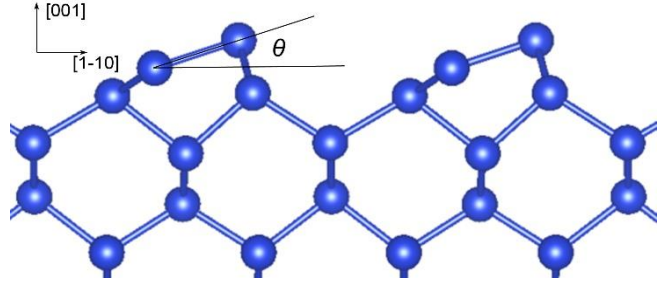

FIG. 7. Geometric structure of the 2×1 reconstructed surface of the Si(001). $\theta$ is the tilting angle of the surface dimers.

The structure of the Si(001) surface has been intensively studied due to its importance both fundamentally in surface science and epitaxial growth [20-23] and practically in fabrication of devices [68]. Fig. 7 shows a 2×1 reconstructed structure on Si(001) surface that spontaneously forms from the as-cut surface. In this 2×1 reconstruction, surface atoms dimerize to form additional bonds. Because of the Jahn-Teller effect, the surface dimer tilts by an angle $\theta$ of 18° to break the degeneracy of surface band [69,70], in order to lower the surface energy. This electronic effect has not been captured by any empirical potentials. All the empirical potentials give rise to $\theta$=0° (except $\theta$=4° for EDIP), which is too small [48]. GAP and DFTB can do a slightly better job to achieve an angle with about -2.5° and -2.3° errors, respectively [48]. In Fig. 8, we show the formation energy of the reconstruction of the Si(001) surface with different reconstructed structures (2×1, 2×2, 4×1 and 4×2). The surface energies calculated by our NN potential match well with the DFT reference. The calculated tilting angle $\theta$=19.4°, which is 1.4° larger than the one obtained from the DFT results, representing the best among all the non-DFT potentials.

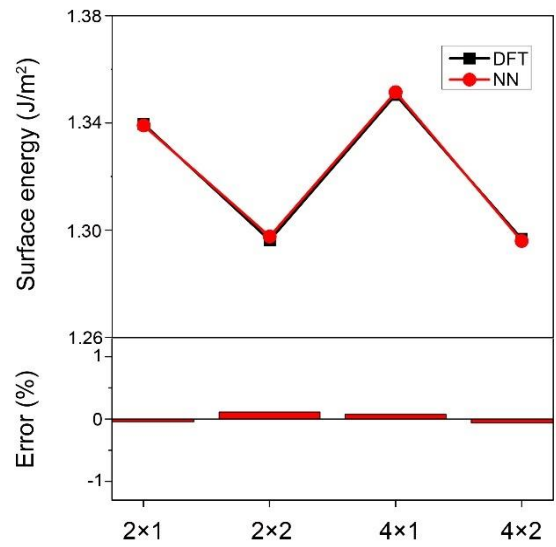

FIG. 8. Formation energies of the Si(001) reconstructed surfaces for various surface unit cell sizes 2×1, 2×2, 4×1 and 4×2, computed with NN potential (red line) and DFT (black line). Comparison of percentage errors are made with respect to DFT reference.

## 2. Si(111) reconstructed surface

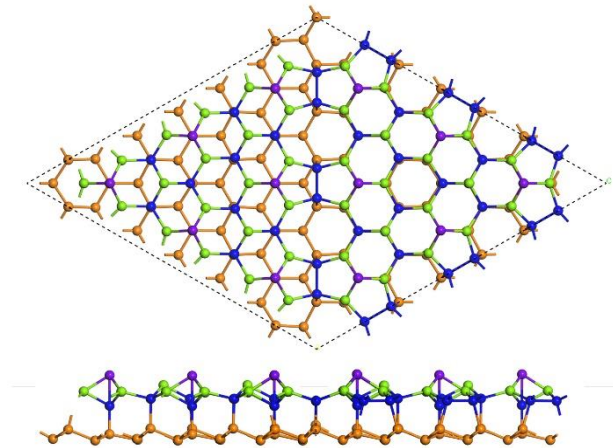

FIG. 9. Top and side view of geometric structure of 7×7 reconstruction in Si(111) surface. The purple, green, blue and brown color balls represent first (adatoms), second, third and fourth layer, respectively.

It is known that 7×7 reconstruction has the lowest energy in Si(111) surface. Since the first observation of 7×7 reconstruction on Si(111) surface in 1959 [71], the atomic structure of 7×7 reconstruction has been the subject of intensive attention. Although many models were proposed for the 7×7 reconstruction, only the one proposed by Takayanagi *et al.* [72], agreed with all the experimental measurements. The dimer-atomic-stack-fault (DAS) model has a fairly complex structure involving a 2D superlattice of polyatomic rings, as shown in Fig. 9. There are dimerization dislocations connecting the core. The stabilization of this model is mainly due to the balance of charge transfer and stress. Other DAS-type reconstructions may also be obtained under special conditions, such as rapid quenching from disordered 1×1 structure [73].

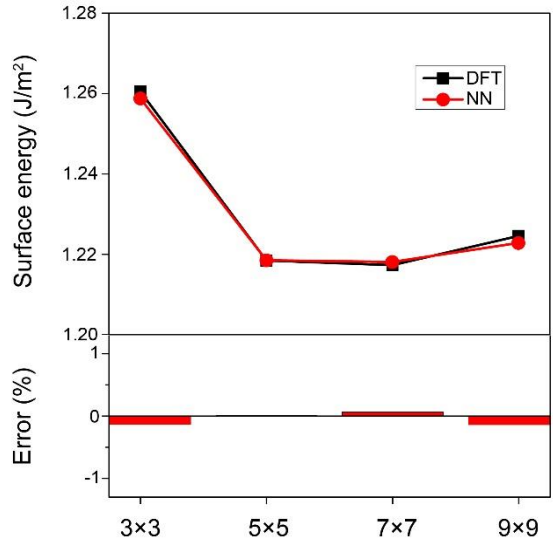

FIG. 10. Formation energies of the Si(111) DAS reconstructed surfaces for various surface unit cell sizes 3×3, 5×5, 7×7 and 9×9, computed with NN potential (red line) and DFT (black line). Comparison of percentage errors are made with respect to DFT reference.

Due to the complex structures and charge transfer involved, all of the empirical potentials, as well as DFTB method, cannot successfully simulate this $7 \times 7$ reconstruction [48]. The recently developed GAP method, with an error below 0.05 $J/m^2$, correctly predicts the DAS family to be lower in energy than the unreconstructed surface. However, GAP method wrongly predicts the 5×5 reconstruction as the lowest reconstruction, 0.01 J/m$^2$ lower than the 7×7 reconstruction [48]. As shown in Fig. 10, the surface energies predicted by our NN potential agree perfectly with the DFT results. For the 3×3, 5×5 and 7×7 reconstructions, the errors are all smaller than 0.001 J/m$^2$, and the error of 9×9 reconstruction is a little larger, ~0.0017 J/m$^2$. Most importantly, the of 7×7 reconstruction is predicted to be the ground-state by our NN potential. It is noted that the 7×7 and 9×9 reconstructed structures are not included in the training database, yet our NN potential still gives a good quantative prediction. This indicates our NN potential not only has a very good transferability, but also shows a surprising "extrapolation" ability! We also used our NN potential to do the optimization of 7×7 reconstruction with quasi-Newton optimized method in Atomic Simulation Environment [74]. As shown in Table 2, we have compared several critical structural parameters in the DAS reconstructions with the DFT reference. We found that the structural parameters agree well between the two methods, especially for the parameters related to the dimer chain. Therefore, our NN potential provides the first general interatomic potential that can give a very good description and prediction of the DAS reconstruction in Si(111) surface.

TABLE 2. Structural parameters in the DAS reconstruction: average height of the adatoms from the atoms directly below in the third-layer $d_1$, average bond length of the adatoms to the three atoms directly bonded in the second-layer $d_2$, average distance between the atoms which directly bond with adatoms in the second-layer $d_3$, average bond length of dimer chain in the third-layer $d_4$, average bond length of the atoms in the dimer with the atoms in the forth-layer $d_5$, and the average bond length of the atoms in the dimer with the atoms in the second-layer $d_6$.

| Model | $d_1$(Å) | $d_2$(Å) | $d_3$(Å) | $d_4$(Å) | $d_5$(Å) | $d_6$(Å) |
|---|---|---|---|---|---|---|
| DFT | 2.45 | 2.48 | 3.71 | 2.44 | 2.40 | 2.40 |
| NN potential | 2.42 | 2.58 | 3.85 | 2.46 | 2.37 | 2.39 |

## IV. CONCLUSION

We have developed a generalized ML-based NN potential for Si, which accurately reproduces DFT reference results for a wide range of physical properties of different Si systems, including bulk crystal, liquid, amorphous and surface. Especially, our NN potential is the first and only one to date that can give an excellent description of the tilted Si dimers in the Si(001) surface and the DAS reconstruction in the Si(111) surface. We envision that our accurate NN Si potential will pave the way to many future studies of complex Si systems, such as simulation of growth and nucleation processes in Si that require large system size and long time scale, which were intractable before. Our approach of developing this accurate and transferable Si NN potential can be generally to other materials.

**Note:** During the preparation of our manuscript, we notice an independent paper on discussing ML-based Si potential published [48]. There are *pros and cons* of the two potentials, and our potential is especially developed for more accurate description of Si surface properties, e.g., correctly predicting the ground-state of Si(111)-7x7 reconstruction.

**Acknowledgements**

We thank Prof. Suhuai Wei, Dr. Bin Cui and Dr. Ninghai Su for help discussions. L.H. and B.H. acknowledge the support from Science Challenge Project (No. TZ2016003), China Postdoctoral Science Foundation (No. 2017M610754), NSFC (Grant No. 11574024 and No. 11704021) and NSAF (No. U1530401). F. L. acknowledges the support from US-DOE (Grant No. DE-FG02-04ER46148). Computations were performed at Tianhe2-JK at CSRC and Paratera Clusters.

**References**

1. M. W. Finnis, Interatomic Forces in Condensed Matter (Oxford University Press, 2004).

2. M.W. Finnis, J.E. Sinclair, Philos. Mag. A 50 (1984) 45–55.

3. F.H. Stillinger, T.A. Weber, Phys. Rev. B 31 (1985) 5262–5271.

4. J. Tersoff, Phys. Rev. B 37 (1988) 6991–7000.

5. D.W. Brenner, Phys. Rev. B 42 (1990) 9458–9471.

6. M.S. Daw, S.M. Foiles, M.I. Baskes, Mater. Sci. Rep. 9 (1993) 251–310.

7. S.J. Stuart, A.B. Tutein, J.A. Harrison, J. Chem. Phys. 112 (2000) 6472.

8. C. M. Bishop, Pattern Recognition and Machine Learning (Springer, 2016).

9. J. Behler, M. Parrinello, Phys. Rev. Lett. 98 (2007) 146401.

10. J. Behler, Int. J. Quant. Chem. 115 (2015) 1032–1050.

11. A.P. Bartók, M.C. Payne, R. Kondor, G. Csányi, Phys. Rev. Lett. 104 (2010) 136403.

12. A.P. Bartók, G. Csányi, Int. J. Quantum Chem. 115 (2015) 1051–1057.

13. J. Behler, J. Chem. Phys. 145, 170901 (2016).

14. R. Ramprasad, R. Batra, G. Pilania, A. Mannodi-Kanakkithodi, and C. Kim, npj Computational Materials, 3, 54 (2017).

15. D. R. Hamann, M. Schlüter, and C. Chiang, Phys. Rev. Lett. 43, 1494 (1979).

16. T.A. Arias and J.D. Joannopoulos, Phys. Rev. Lett. 73, 680 (1994).

17. F. Liu, M. Mostoller, V. Milman, M.F. Chisholm, and T. Kaplan, Phys. Rev. B 51, 17192 (1995).

18. G. Binnig, H. Rohrer, Ch. Gerber, and E. Weibel, Phys. Rev. Lett. 50, 120 (1983).

19. B. S. Swartzentruber, Y.-W. Mo, R. Kariotis, M. G. Lagally, M. B. Webb, Phys. Rev. Lett. 65, 1913 (1990).

20. V. Zielasek, Feng Liu, Yuegang Zhao, J.B. Maxson, and M.G. Lagally, Phys. Rev. B, 64, R201320 (2001).

21. F.-K Men, F. Liu, P.J. Wang, C.H. Chen, D.L. Cheng, J.L. Lin, and F.J. Himpsel, Phys. Rev. Lett. 88, 096105 (2002).

22. "Epitaxial Growth of Si on Si(001)", Feng Liu and M.G. Lagally, in "The
23